\documentclass[aps,showpacs,groupedaddress,floatfix]{revtex4}
\usepackage{amsmath,epsfig,bm,pifont}

\usepackage{graphicx} 
\usepackage{bm} 
\usepackage{amssymb}
\usepackage{dcolumn}
\usepackage{amsmath}
\usepackage[normalem]{ulem}

\begin{document}

\def\gsim{\mathrel{\rlap{\lower4pt\hbox{\hskip1pt$\sim$}}}}
\newcommand{\beq}{\begin{equation}}
\newcommand{\eeq}{\end{equation}}
\newcommand{\beqn}{\begin{eqnarray}}
\newcommand{\eeqn}{\end{eqnarray}}
\newcommand{\btab}{\begin{tabular}}
\newcommand{\etab}{\end{tabular}}
\newcommand{\To}{\Longrightarrow}
\newcommand{\brho}{\mbox{\boldmath$\rho$}}
\newcommand{\bome}{\mbox{\boldmath$q_0$}}
\newcommand{\re}{\nonumber\\}
\newcommand{\etal}{{\em{et al.}}}
\newcommand{\ibid}{{\em{ibid}}}
\newcommand{\tm}{\times}
\newcommand{\lc}{\left<}
\newcommand{\rc}{\right>}
\newcommand{\lr}{\left|}
\newcommand{\rl}{\right|}
\newcommand{\lb}{\left(}
\newcommand{\rb}{\right)}
\newcommand{\ls}{\left[}
\newcommand{\rs}{\right]}
\newcommand{\Lb}{\left\{}
\newcommand{\Rb}{\right\}}
\newcommand{\xchi}{X_\sigma}
\newcommand{\wxchi}{\widetilde{X}_\sigma}
\def\ms{M_s}
\def\mst{\tilde{M}_s}
\newcommand{\qq}{{\bf q}}
\newcommand{\kf}{k_{\rm F}}
\newcommand{\pp}{{\bf p}}
\newcommand{\kk}{{\bf k}}
\newcommand{\hh}{{\bf h}}
\newcommand{\tlq}{\tilde{q}}
\newcommand{\tlt}{\tilde{T}}
\def\gsim{\mathrel{\rlap{\raise 0.5ex
     \hbox{$>$}}{\lower 0.55ex \hbox{$\sim$}}}}
\def\lsim{\mathrel{\rlap{\raise 0.5ex
      \hbox{$<$}}{\lower 0.55ex \hbox{$\sim$}}}}
\newcommand{\gcq}{\, \text{g}/\text{cm}^{3}}
\newcommand{\la}{\langle}
\newcommand{\ra}{\rangle}
\newcommand{\ts}{\textstyle}
\newcommand{\wt}{\widetilde}

{\begin{flushright} 
INT-PUB-13-013
\vskip -0.5in
{\normalsize }
\end{flushright}

\title{Neutrino scattering from hydrodynamic modes in hot and dense neutron matter}

\author{Gang Shen}
\affiliation{Institute for Nuclear Theory, University of Washington, 
Seattle, Washington 98195, USA}
\author{Sanjay Reddy}
\affiliation{Institute for Nuclear Theory, University of Washington, 
Seattle, Washington 98195, USA}

\begin{abstract}
We calculate the scattering rate of low energy neutrinos in hot and dense neutron matter encountered in neutrons stars and supernova in the hydrodynamic regime. We find that the Brillouin peak, associated with the sound mode, and the Rayleigh peak, associated with the thermal diffusion mode, dominate the dynamic structure factor. Although the total scattering cross section is constrained by the compressibility sum rule, the differential cross-section calculated using the hydrodynamic response function differs from results obtained in approximate treatments often used in astrophysics such as random phase approximations (RPA). We identified these differences and discuss its implications for neutrino transport in supernova.        
\end{abstract}
\date{\today}
\maketitle

The energy spectrum of neutrinos emitted from core collapse supernova plays a crucial role in several aspects of supernova dynamics, neutrino oscillations, supernova nucleosynthesis, and their detectability in terrestrial neutrino detectors. The spectrum is determined by neutrino interactions in the outer layers of the proto-neutron star (PNS) called the neutrino-sphere. Neutron-rich matter encountered in the neutrino-sphere have densities and temperatures in the range $10^{12}-10^{14}$ g/cm$^3$, and $T=3-8$ MeV, respectively. Under theses conditions, the neutrino-nucleon scattering rate is modified by strong interactions between nucleons\cite{Sawyer:1975,Iwamoto:1982}.  

Neutrino scattering off non-relativistic nucleons in dense matter can be related to the density-density and spin-density nucleon correlation functions of the hot and dense nuclear plasma \cite{Iwamoto:1982}. In this study we only consider scattering off density fluctuations as our interest is to understand specific aspects of the long-time response which we discuss in more detail below.  In this case, the differential cross section for the neutral current reaction $\nu N \rightarrow \nu N$ is given by  
\beq
\frac{1}{V}\frac{d^2\sigma(E)}{d{\rm cos}\theta dE'}=\frac{G_F^2{\rm cos}\theta^2}{4\pi^2}c_v^2\bigl[1+{\rm cos}\theta\bigr]
E'^2[1-f(E')] S(q,q_0)
\label{Eq.sigma}
\eeq
where $S(q, q_0)$ called the dynamic structure factor is the quantity of interest and describes the response of the strongly interacting neutron gas. The other symbols that appear in the above equation are: $G_F$ is the Fermi constant, the neutron weak vector charge is $c_v=0.5$, $\theta$ the scattering angle, $E$ and $E'$ are the initial and final neutrino energies and $f(E')$ the final state neutrino blocking factor. Typically $S(q,q_0)$ within the framework of Landau's quasi-particle picture, and is justified when $q_0\tau \gg 1$, where $q_0$ is the characteristic energy transfer to the nucleonic systems during the scattering process and $\tau$ is the lifetime of quasi-particle.   The residual interactions between these quasi-particles are included by diagrammatic re-summation techniques such as the random-phase-approximation (RPA) which incorporate long-range correlations in the one-(quasi)particle-hole excitations \cite{Horowitz91,Reddy:1998,Reddy:1999,Burrows98,Burrows99}.

In the opposite limit, when $q_0\tau \lsim 1$, the response is characterized by the long-time behavior of the system and multiple collisions between nucleons become relevant. Here, it is well-known that hydrodynamics provides an accurate description of the density-density response function \cite{hydro}. Motivated by the observation that for a wide range of ambient conditions in the neutrino-sphere region and for typical thermal neutrino energies, $q_0\tau \lsim 1$, we have calculated the density-density response function in the hydrodynamic limit and compared our results with earlier results obtained in the quasi-particle picture. 

Using a moment expansion for solving the linearized Boltzmann equation in hydrodynamic regime \cite{Watabe:2009,Watabe:2010}, we have calculated  the response function relevant to neutrino scattering which includes collective modes as well as the hydrodynamic response function in hot and dense neutron matter. We find that the total scattering cross section is well constrained by the compressibility sum rule from the underlying equation of state.  Thus approximate methods such as Random Phase Approximation (RPA) which satisfy the compressibility sum rule can be used to calculate the total cross-section. However, we find that the differential cross-section obtained the hydrodynamic approach differs in several respects from that obtained in RPA and may have implications for neutrino transport. 

We assume that the interactions between neutrons can be approximated by an effective zero range force with a $S$--wave scattering strength controlled by the dimensional parameter $g$. This should be a good approximation to the low density neutron rich matter we are interested in for neutrino-sphere in PNS \cite{Gezerlis:2009}. In our approach we fix $g$ by calculating the density response function in a specific approximation and then matching to the compressibility obtained in the low density limit for a realistic equation of state. Once $g$ is determined in this way, the collective modes and dynamic response function in hydrodynamic regime can be calculated for the normal fermi gas as described in some detail in Refs.~\cite{Watabe:2009,Watabe:2010}. Here we simply note their main result which states that in the moment approach the hydrodynamic equations for oscillations of density ($\delta n_{\rm tot}$), momentum (${\bf q}\cdot \overline{\bf v}$), and energy ($\delta E$) in systems under external perturbation ($U_{\rm ext}$) can be expressed in a matrix form: 
\begin{align}
M
\begin{pmatrix}
\delta n_{\rm tot}  \\
{\bf q}\cdot \overline{\bf v} \\
\delta E 
\end{pmatrix} 
\equiv 
\begin{pmatrix}
q_0 & \displaystyle{\frac{2F_{2}}{3m}} & 0 \\
-\displaystyle{\frac{gq^{2}}{2m}} & \displaystyle{q_0 - i\frac{2\eta q^{2}}{F_{2}} }& \displaystyle{\frac{q^{2}}{F_{2}}} \\
\displaystyle{-i\frac{\Gamma_{\kappa} \gamma F_{2}}{2m F_{0}} }& \displaystyle{\frac{F_{4}}{3m^{2}} }& q_0 + i \Gamma_{\kappa} \gamma
\end{pmatrix}
\begin{pmatrix}
\delta n_{\rm tot}  \\
{\bf q}\cdot \overline{\bf v} \\
\delta E 
\end{pmatrix} 
= 
\begin{pmatrix}
0 \\
 \displaystyle{\frac{q^{2}}{m}  U_{\rm ext}  }\\
0
\end{pmatrix}, 
\label{Matrix14}
\end{align}
where $({\bf{q}},q_0)$ is four momentum transfer, $m$ is mass of nucleon, and $F_n$ is defined as follows, 
\beq 
F_n\ =\ \int \frac{d^3p}{(2\pi)^3} \frac{\partial f_0}{\partial \epsilon_0} p^n,
\eeq
where $f_0=1/(1+e^{\beta(\epsilon_0-\mu)})$ is Fermi-Dirac function for free fermi gas with $\beta$ the inverse temperature, $\mu$ the chemical potential, and $\epsilon_0$ the energy of free nucleon. In Ref.~\cite{Benhar:2010} the shear viscosity $\eta$ and thermal conductivity $\kappa$ has been calculated by solving the Boltzmann equation for neutrons. Although we will use their results we briefly mention as an aside that a simple estimate can be made using kinetic theory these can be written as  
\beqn
\eta\ =\ \frac 1 3 n \bar{p} \lambda, \ \ \kappa \ =\ \frac 1 3 C_V \bar{p} \lambda /m,
\eeqn
where $n$ is number density, $\bar{p}$ is average momentum, $C_V$ is specific heat $\lambda \approx 1/n_n \sigma_n$ where $n_n$ is the neutron density and $\sigma_n$ is the neutron-neutron scattering cross-section including Pauli blocking. We have found that this simple estimate agrees fairly well (within a factor of 2) with the results from Ref.~\cite{Benhar:2010}. We also note that the transport coefficients don't influence the matching between the static structure factor and compressibility from equation of state, since in the long wavelength limit ($q\rightarrow 0$) the transport coefficients drop out. However, as we discuss later they determine the width of the collective modes and the shape of the response functions. 

Solving the matrix equation (\ref{Matrix14}) for $\delta n_{\rm tot}$, 
one can obtain 
\begin{align}
\delta n_{\rm tot} ({\bf q}, q_0)
&  = - \frac{2F_{2}}{3m^{2}}q^{2} \frac{q_0 + i\Gamma_{\kappa} \gamma}{{\rm det} M} U_{\rm ext}({\bf q}, q_0)
\label{eq:dn}
\end{align} 
where
\begin{align}
\gamma \equiv \frac{F_{0}(F_{4}-gF_{2}^{2})}{F_{2}^{2}(1-gF_{0})} \,, {\rm{and}}\ \
\Gamma_{\kappa} \equiv - \frac{2 \kappa T m^{2} q^{2} F_{2}^{2} (1-g F_{0})}{(F_{4}-g F_{2}^{2})(F_{4}F_{0}-F_{2}^{2})} \,.
\label{Gammak}
\end{align} 
Since $\delta n_{\rm tot} ({\bf q}, q_0)= \chi ({\bf q}, q_0) U_{\rm ext} ({\bf q}, q_0)$ the density response function $\chi ({\bf q}, q_0) $ can be read off easily from equation \ref{eq:dn} and the dynamic structure factor is obtained via fluctuation -dissipation theorem, 
\beq
S(q,q_0)\ =\ -\frac{1}{n\pi} \frac{1}{1-e^{-\beta q_0}} \rm{Im} \chi(q,q_0).
\label{fluc}
\eeq
%
If the second-order terms in the transport coefficients $\kappa$ and $\eta$ are small in the hydrodynamic regime, the determinant of the matrix $M$ can be reduced to 
${\rm det} M = (q_0^{2} - \Omega^{2}) ( q_0 + i \Gamma_{\kappa} ) + 2 i \Gamma q_0^{2}$, 
where 
\begin{align}
\Omega
\equiv  
\sqrt{\frac{F_{4}-g F_{2}^{2}}{3 F_{2}}}\frac{q}{m} \equiv c q, {\rm{and}} \ \ 
\Gamma = - \frac{\eta q^{2}}{F_{2}} - \frac{\kappa T q^{2} m^{2}}{(F_{4}-g F_{2}^{2})}. 
\label{ReFirstSound}
\end{align}
The poles of the determinant gives the eigenmodes of the hydrodynamic modes (to first order in $\kappa$ and $\eta$): 
$q_0 = \pm \Omega - i \Gamma$ and $q_0 = -i\Gamma_{\kappa}$~\cite{Watabe:2009}. 
$c$ in Eq.~(\ref{ReFirstSound}) is the sound velocity. 
$\Gamma$ and $\Gamma_{\kappa}$ are damping rates of sound mode and Rayleigh diffusion mode, respectively. 

Based on above approximations, one can explicitly derive the resultant density response function and the absorptive susceptibility, imaginary part of density response function, follows  
\begin{align}
{\rm Im}\chi ({\bf q}, q_0) 
= & \frac{2 F_{2}}{3m^{2}c^{2}} 
\left [ 
\frac{q_0 (\gamma-1)\Gamma_{\kappa}}{q_0^{2}+\Gamma_{\kappa}^{2}}
+ 
\frac{2 q_0 \Gamma\Omega^{2}}{(q_0^{2}- \Omega^{2})^{2} + (2 q_0 \Gamma)^{2}}
\right . 
\nonumber
\\
&
\left .
-
\frac{q_0\Gamma_{\kappa} (\gamma-1)(q_0^{2}-\Omega^{2})}{(q_0^{2}-\Omega^{2})^{2} + (2 q_0 \Gamma)^{2}}
\right ], 
\label{smallq}
\end{align} 
where $\Omega \gg \Gamma_{\kappa}$, and also $\Omega \gg \Gamma$ were used. 
The absorptive susceptibility ${\rm Im}\chi ({\bf q}, q_0)$  has two peaks: 
the Rayleigh diffusion peak at $q_0 = 0$ and the Brillouin peak at $q_0 = \Omega$. 
  
The static structure function
\beq
S(q)\ =\ \int_{-\infty}^{\infty} S(q,q_0) d q_0
\eeq
is related to the compressibility of matter  $(\partial P/\partial n)_T$ in the long wavelength limit through the  compressibility sum-rule
\beq
S(q\rightarrow 0) \ = \ -\frac{2T F_2}{3 nm^2 c^2}\gamma \equiv \frac{T}{(\partial P/\partial n)_T},
\eeq
We use this relation to determine the parameter $g$ from the compressibility of underlying equation of state, and for the results we present in this study we use the NL3 nuclear equation of state \cite{Shen:2011}. We note that the hydrodynamic responses below are obtained from Eqs. (\ref{eq:dn}, \ref{fluc}) numerically, not from approximate Eq. (\ref{smallq}).

\begin{figure}[h]
\includegraphics[width=0.45\columnwidth]{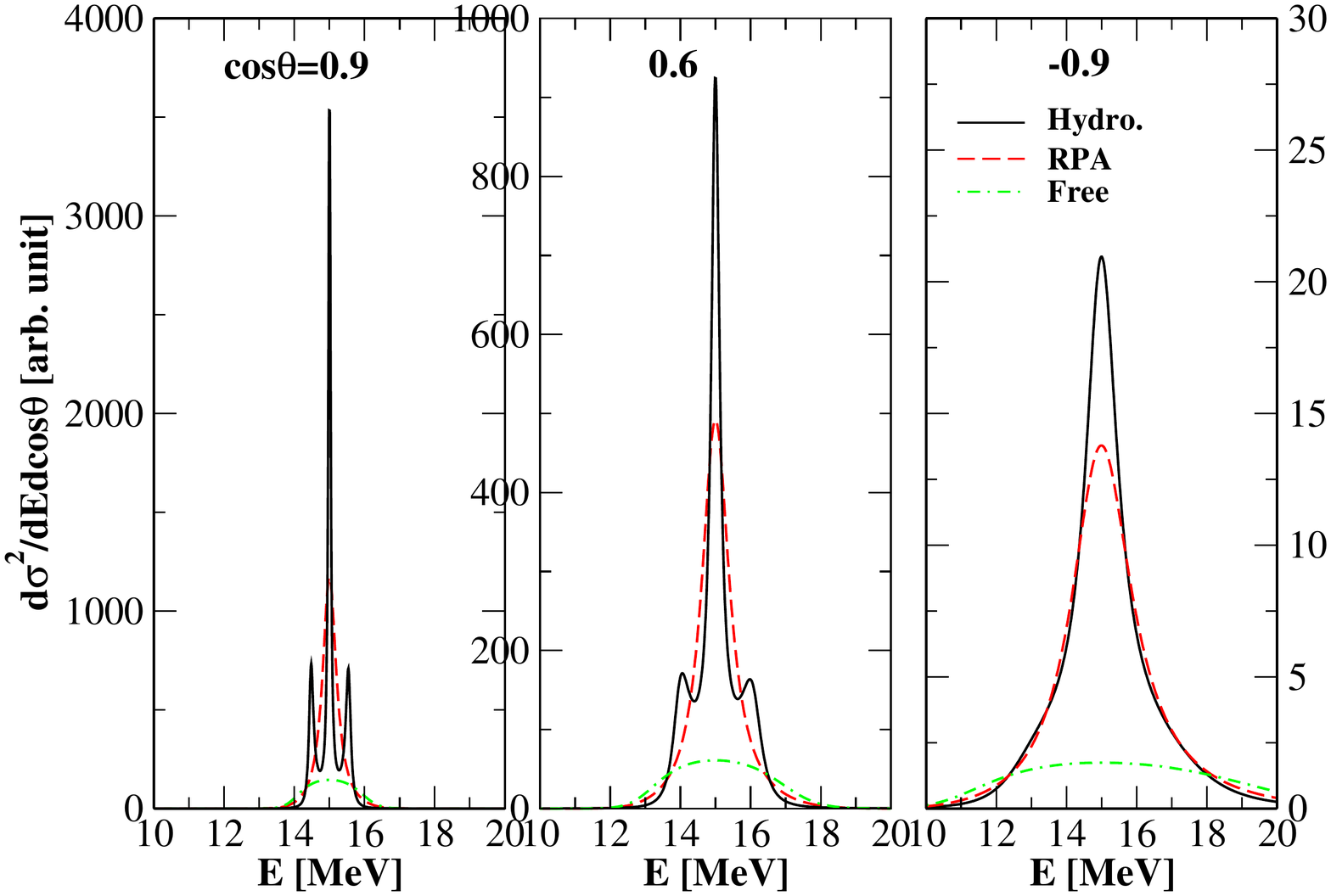}
\includegraphics[width=0.45\columnwidth]{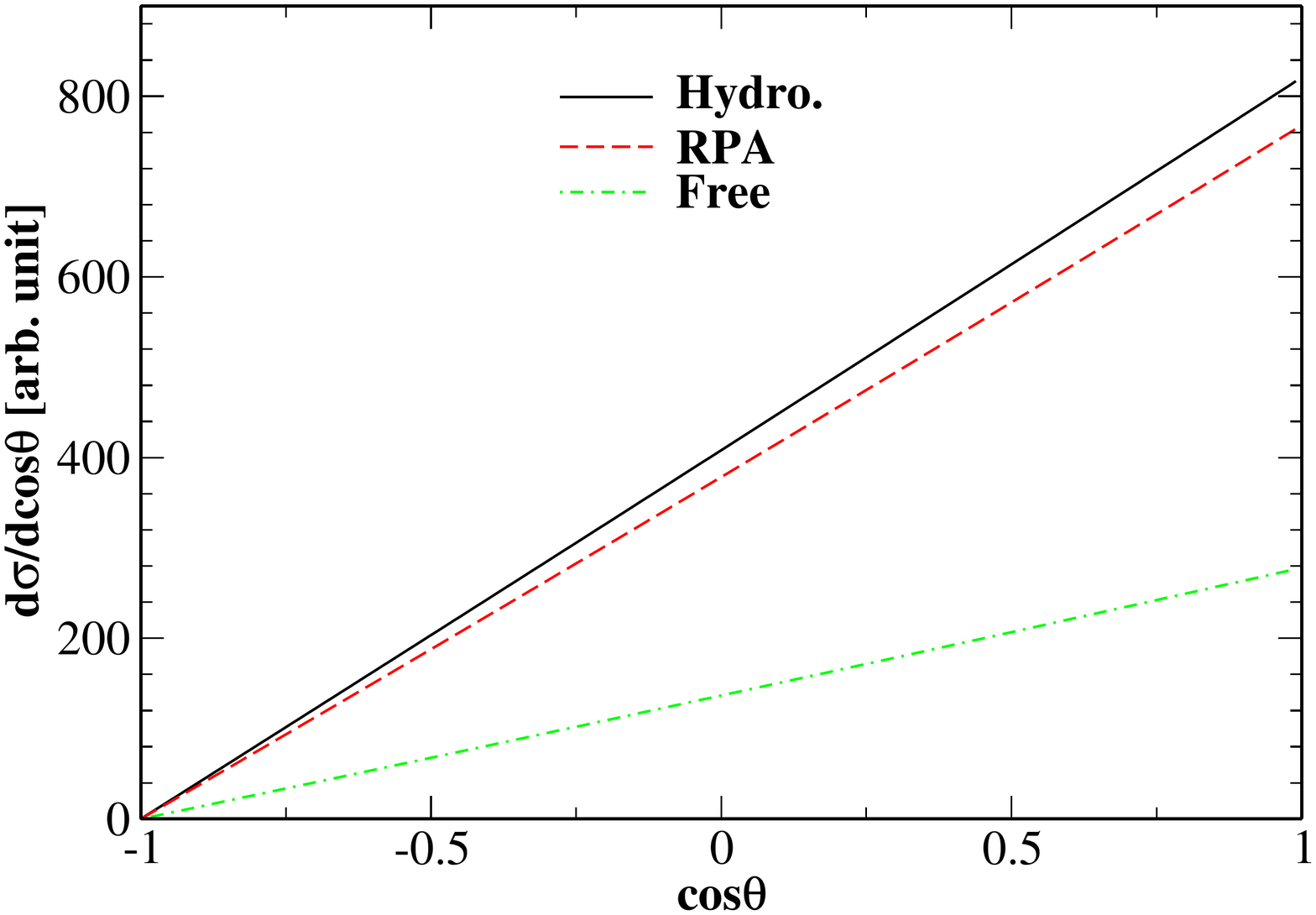}
\caption{(Color online) Differential cross section versus final lepton energy (left Panel) for various scattering angle, $\mathrm{cos\theta}$ = 0.9, 0.6, and --0.9, and differential cross-section versus $\cos{\theta}$ (right panel) obtained in RPA, hydrodynamic response, and free fermi gas. Neutron density $n$= 10$^{-2}$ fm$^{-3}$, temperature $T$ = 5 MeV and incident neutrino energy is $E_\nu=3 T$.}
 \label{fig:1}
\end{figure}

Fig.~\ref{fig:1} (left panel) shows the differential cross section versus final neutrino energy for various scattering angle, $\mathrm{cos\theta}$ = 0.9, 0.6, and --0.9, obtained in RPA, hydrodynamic response, and free fermi gas. Neutron matter is at $n$= 10$^{-2}$ fm$^{-3}$ and $T$ = 5 MeV. The incident neutrino energy is 3$T$. In RPA, the residual quasi-particle-hole interaction is derived from derivative of potential energy, which satisfies static sum rules similar to hydrodynamic response. The force is attractive at this density and enhances response of nucleon gas compared to free fermi gas in each angle. In this calculation we used NL3 nuclear effective interaction to calculate RPA response (as well as mean field response) for consistency. At forward angle (where momentum transfer is small), the differential cross section from hydrodynamic response clearly exhibits the features of collective modes - the central peak is due to Rayleigh mode and the two side peaks are due to Brillouin mode. At backward angle, the damping to the collective modes becomes so large that the latter differential cross section becomes similar to the one from RPA. The dependences of response function on the scattering angle may influence the neutrino transport in the low density region, particularly for low energy, forward-scattering neutrinos. It would be interesting to study its effect on the spectra of supernova neutrinos in a more detailed simulation. In the right panel of Fig.~\ref{fig:1} we show the differential cross section versus scattering angle, $\rm{cos}\theta$ (after integrating over final lepton energy), obtained in RPA, hydrodynamic response, and free fermi gas. Neutron matter is at $n$= 10$^{-2}$ fm$^{-3}$ and $T$ = 5 MeV. The incident neutrino energy is 3$T$. The angular distribution of the RPA and the hydrodynamic responses are very close to each other and the integration over the final neutrino energies washes out the larger differences seen in the double differential cross-section in the left panel.

%

Table.~\ref{tab:1} shows the total cross section, obtained in mean field (Hartree) approximation, RPA response, hydrodynamic response, and free fermi gas. Neutron matter is at $n$= 10$^{-2}$ fm$^{-3}$ and $T$ = 5 MeV. The NL3 EOS is used to obtain the compressibility. The incident neutrino energy is 3$T$. The neutrino scattering cross section is proportional to $S(q=0)$. The ratio of hydrodynamic response/free fermi gas response is equal to that of quasi-particle RPA/quasi-particle mean field (Hartree). This clearly demonstrates that the compressibility from underlying equation of state strongly constrains the response of medium, whether in the hydrodynamic picture or quasi-particle picture.
\begin{table}[htbp]
   \centering
   \caption{Total cross section per volume $\sigma/V$ in unit of $10^{-4} ~m^{-1}$, obtained using the Fermi gas (FG), Hartree, RPA, and hydrodynamic response functions. Ambient conditions: $n$= 10$^{-2}$ fm$^{-3}$, $T$ = 5 MeV and $E_\nu=3T$. The NL3 EOS is used to obtain the compressibility.} 
   \begin{tabular}{@{} ccccccc @{}} 
    \hline
   FG &  Hartree & RPA & Hydro &  RPA/Hartree & Hydro/FG \\ \hline
       2.97 & 2.75 & 8.20 & 8.82 &  2.98  & 2.97 \\  
\hline
   \end{tabular}
      \label{tab:1}
\end{table}

\begin{figure}[h]
\includegraphics[width=0.7\columnwidth]{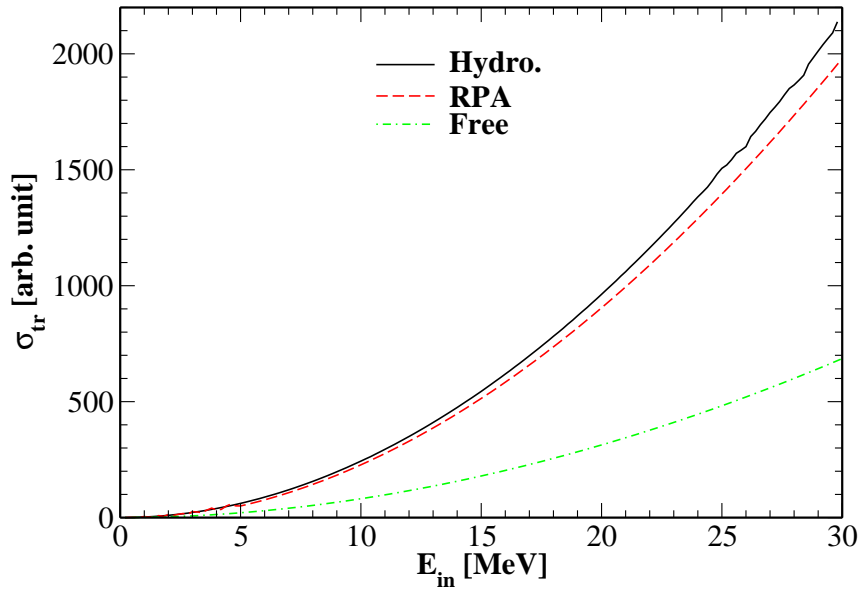}
\caption{(Color online) Transport cross section versus incident neutrino energy, obtained in RPA, hydrodynamic response, and free fermi gas. Neutron matter is at $n$= 10$^{-2}$ fm$^{-3}$ and $T$ = 5 MeV. }
 \label{fig:2}
\end{figure}

Figure \ref{fig:2} shows the total transport cross section versus incident neutrino energy, obtained in RPA, hydrodynamic response, and free fermi gas, as the same conditions in earlier figures. The transport cross section which enters the diffusion equation is weighted by the scattering angle and is defined as  
\beq
\sigma_{tr}\ =\ \int \frac{d\sigma}{d\rm{cos}\theta} (1-\rm{cos}\theta) d\theta.
\eeq
in the elastic limit.  The ratio of hydro./free fermi gas transport cross section is about 3 for almost all incident energy (except when $E_{\nu} \le 1$ MeV), and is close to that of quasi-particle RPA/free Fermi gas as discussed in Table \ref{tab:1}.

In this work we obtained the collective modes and hydrodynamic response of hot and dense neutron matter in the hydrodynamic regime. We found the Brillouin peak in the dynamic structure factor, which is associated with the sound mode, and the Rayleigh peak, which is associated with the thermal diffusion mode. 
We also compared the collisional hydrodynamic response to the collision-less response function based on random-phase-approximation. We find that both yield a very similar result for the total neutrino transport cross section and that its tightly constrained by the compressibility sum rule. At the relatively low densities encountered in the neutrino-sphere attractive nuclear interactions dominate and the iso-thermal compressibility $(\partial P/\partial n)_T$ is reduced, and this in turn enhances the density response increasing the cross-sections by about a factor of 3 for thermal neutrinos.  This will clearly have an impact on the neutrino decoupling temperature and the neutrino spectrum and should be incorporated into supernova simulations. Although the total cross sections are similar, there are differences in the strength distribution between the hydrodynamic and the RPA response. At forward angles, corresponding to modest values of $q$, the sound mode is not strongly damped and appears as bump in the differential cross-section. It would be interesting to  explore if this feature, which enhances energy exchange, can affect the neutrino spectrum formation.



\end{document}